\documentstyle[12pt,epsfig]{article}
\begin{document}
\noindent
\begin{center}
{\Large {\bf Cosmic Evolution in a Modified Brans-Dicke Theory}}\\ \vspace{2cm}
 ${\bf Yousef~Bisabr}$\footnote{e-mail:~y-bisabr@srttu.edu.}\\
\vspace{.5cm} {\small{Department of Physics, Shahid Rajaee Teacher
Training University,
Lavizan, Tehran 16788, Iran}}\\
\end{center}
\vspace{1cm}
\begin{abstract}
We consider Brans-Dicke theory with a self-interacting potential in Einstein conformal frame.  We introduce
a class of solutions in which an accelerating expansion
is possible in a spatially flat universe for positive and large values of the Brans-Dicke parameter consistent with local gravity experiments.
In this Einstein frame formulation, the theory appears as an interacting quintessence model in which the interaction term is given
by the conformal transformation.  In such an interacting model, we shall show that the solutions lead simultaneously to a constant ratio of energy
densities of matter
and the scalar field.
\end{abstract}
\vspace{1cm}
\vspace{3cm}
\section{Introduction}
Cosmological observations
on expansion history of the universe indicate
that the universe is in a phase of accelerated expansion
(Riess et al. 1998 ; Perlmutter et al. 1997a,b).  This phenomenon may be interpreted as evidence either for
existence of some exotic matter components or for modification of
the gravitational theory.  In the first route of interpretation
one can take a mysterious cosmic fluid with
sufficiently large and negative pressure, dubbed dark energy.  This viewpoint also includes quintessence models (Ratra and Peebles 1988; Caldwell, Dave and Steinhardt 1998; Steinhardt, Wang and Zlatev 1999; Macorra and Picinelli 2000; Gonzalez-Diaz 2000; Bludman and Roos 2002; Cai, Saridakis, Setare and  Xia 2010),
which invokes a scalar field minimally coupled to gravity.  The scalar field takes negative pressure during its
evolution by rolling down a proper
potential.  In the second route, however, one attributes the
accelerating expansion to a modification of general relativity.
This includes scalar-tensor theories, scalar fields non-minimally coupled to gravity  (Faraoni 1999; Faraoni 2004).\\
Here we shall consider a self-interacting Brans-Dicke (BD) theory (Brans and Dicke 1961) as a prototype of scalar-tensor theories.  The
original motivation of the BD theory was the search for a theory containing
Mach's principle which has found a limited expression in general relativity.  As
the simplest and best-studied generalization of general relativity, it is natural to think
about the BD scalar field as a possible candidate for producing cosmic acceleration without invoking a quintessence field or
 exotic matter systems (Setare 2007; Setare and Jamil 2010).  In fact, there have been many
attempts to show that BD model can potentially
explain the cosmic acceleration.  It is shown that this theory can actually produce a non-decelerating
expansion for low negative values of the BD
 parameter $\omega$ (Banerjee and Pavon 2001). This conflicts with the lower
bound imposed on this parameter by solar system experiments (Will 1993; Will 2005). Some authors
propose modifications of the BD model such as introducing some potential functions for the scalar field (Bertolami and Martins 2000; Mak and Harko 2005), or
considering a field-dependent BD parameter (Chakraborty and Debnath), without resolving this problem.\\
All the works in this context use Jordan frame representation of BD theory.  It is however well-known
that this theory, like any other scalar-tensor theories, can be represented in the so-called Einstein frame
by using a conformal transformation (Faraoni 1999; Faraoni 2004).  Although these two conformal frames are mathematically
equivalent there are some debates on their physical equivalence.  Here there is a point which should be made clear.  Despite the fact that the problem of physical status of the two conformal
frames is open there is a tendency in the literature to ignore this problem
and to work in Jordan conformal frame. The reason may be related to reluctance in accepting of the violation of weak equivalence principle
due to anomalous coupling of the scalar field to matter systems in Einstein frame.  It is however important to keep in mind that
the physical metric should be singled out already in the vacuum sector of the theory and the coupling of a given metric
to matter systems is determined by the physical significance ascribed to it (Magnano and Sokolowski 1994).  Thus a
criterion based on the coupling of matter with gravity would be effective only if the physical frame were
determined on an independent ground.  Apart from this point, anomalous gravitational coupling in Einstein conformal frame does not
necessarily mean violation of weak equivalence principle.  There is still a possibility
that the effective mass of the scalar field be scale dependent. In this chameleon mechanism (Khoury and Weltman 2004 a,b), the scalar field may
acquire a large effective mass in Solar System scale so that it hides
local experiments while at cosmological scales it is effectively light
and can provide an appropriate cosmological behavior. \\\\ Along these lines, we would like to consider Einstein frame formulation of the theory as a representation
which provides different possibilities in a cosmological setting with respect to the Jordan frame.  In particular, we deal with two important issues.
In section 2, we focus on the question that whether it is possible to achieve accelerating
expansion of the universe for sufficiently large values of the parameter $\omega$.  We will show that the answer is affirmative within a class of solutions which
corresponds to a specific form of the potential function of the BD scalar field.  In section 3, we consider the coincidence problem which concerns with
the coincidence between the observed vacuum energy density and the current matter density.  Recently, there is a large amount of interest
to realize this problem as a consequence of an interaction between matter systems and the dark sector (Zimdahl, Pavon and Chimento 2001; Zimdahl and Pavon 2003; Campo, Herrera and Pavon 2009; Bisabr 2010a).  Although the whole idea seems to be promising, however, the
suggested interaction terms are usually phenomenological and are not generated by a fundamental theory.  In the model presented here, there
is an interaction between the BD scalar field and matter systems induced by conformal transformations. We will investigate the consequences of this interaction term
and derive an expression for the relative energy densities.  We also study the evolution of the latter and show that even though the two types of energy densities evolve differently as the universe expands, the ratio can take a small constant configuration at late times.

\section{Accelerating Expansion}
We begin with a modified form of the BD action in Jordan frame
\begin{equation}
S_{JF}= \int d^{4}x \sqrt{-\bar{g}} (\phi \bar{R} -\frac{\omega}{\phi}\bar{g}^{\mu\nu}\bar{\nabla}_{\mu}\phi \bar{\nabla}_{\nu}\phi-V(\phi))
+S_{m}(\bar{g}_{\mu\nu}, \psi)\label{a1}\end{equation}
where $\phi$ is the self-interacting BD scalar field with a potential function $V(\phi)$, $\omega$ is a constant parameter and  $S_{m}$ is the action
of matter which depends on the metric $\bar{g}_{\mu\nu}$ and some matter
fields collectively denoted by $\psi$. A conformal transformation
\begin{equation}
\bar{g}_{\mu\nu}\rightarrow g_{\mu\nu}=\Omega^2 \bar{g}_{\mu\nu}
\label{a2}\end{equation}
with $\Omega=\sqrt{G \phi}$ brings the above action into the Einstein frame (Faraoni 1999; Faraoni 2004).  Then a scalar field redefinition
\begin{equation}
\varphi(\phi)=\sqrt{\frac{2\omega+3}{16\pi G}}\ln (\frac{\phi}{\phi_0})
\label{a3}\end{equation}
with $\phi_0\sim G^{-1}$, $\phi>0$ and $\omega>-\frac{3}{2}$ transforms the kinetic term of the scalar field into a canonical form.  In terms of the variables
($g_{\mu\nu}$, $\varphi$) the BD action in the Einstein frame is (Faraoni 1999; Faraoni 2004)
\begin{equation}
S_{EF}= \int d^{4}x \sqrt{-g} (\frac{R}{16\pi G} -\frac{1}{2}g^{\mu\nu}\nabla_{\mu}\varphi \nabla_{\nu}\varphi-U(\varphi))+S_{m}(g_{\mu\nu}, \psi)
\label{a4}\end{equation}
where
\begin{equation}
S_{m}= \int d^{4}x \sqrt{-g}~ \exp(-8\sqrt{\frac{\pi G}{2\omega+3}}\varphi)~L_{m}(g_{\mu\nu}, \psi)
\label{a5}\end{equation}
Here $\nabla_{\mu}$ is the covariant derivative of the rescaled metric $g_{\mu\nu}$ and
\begin{equation}
U(\varphi)= V(\phi(\varphi))~\exp(-8\sqrt{\frac{\pi G}{2\omega+3}}\varphi)
\label{a6}\end{equation}
is the Einstein frame potential.   \\
Variation of the action (\ref{a4}) with respect to $g_{\mu\nu}$ and $\varphi$ leads to the following field equations
\begin{equation}
G_{\mu\nu}=8\pi G (T_{\mu\nu}+T^{\varphi}_{\mu\nu})
\label{a7}\end{equation}
\begin{equation}
\Box\varphi-\frac{dU(\varphi)}{d\varphi}=-\frac{1}{2}\alpha~T
\label{a8}\end{equation}
where
\begin{equation}
T^{\varphi}_{\mu\nu}=\nabla_{\mu}\varphi \nabla_{\nu}\varphi-\frac{1}{2}g_{\mu\nu}\nabla_{\gamma}\varphi \nabla^{\gamma}\varphi-U(\varphi)g_{\mu\nu}~,
\label{a9}\end{equation}
$\alpha = \sqrt{\frac{16\pi G}{2\omega+3}}$ and $T=g^{\mu\nu}T_{\mu\nu}$ is the trace of the matter stress-tensor.  Note that the parameter
$\alpha$ is related to inverse of the BD parameter $\omega$.  In Einstein frame, the vacuum sector of the action consists of a scalar
field minimally coupled to Einstein's gravity.  The important difference between the Einstein frame representation of BD model and minimally coupled scalar
field models is that in the former the scalar field interacts with matter systems.  This anomalous gravitational coupling has no counterpart in Einstein's gravity.  It implies that the stress-tensors of matter and the scalar field are not
separately conserved.  This can be easily checked by applying Bianchi's identities to (\ref{a7}) which leads to
\begin{equation}
\nabla^{\mu}T_{\mu\nu}=-\nabla^{\mu}T^{\varphi}_{\mu\nu}=\frac{1}{2}\alpha~T~ \nabla_{\nu}\varphi
\label{1a9}\end{equation}
The parameter $\alpha$ measures the strength of the interaction.  Here there are two important points in order :
First, the parameter $\alpha$ is positive $\alpha>0$.  It implies that energy transfer is from scalar field $\varphi$ to matter systems.  This feature is consistent with the second law of thermodynamics (Pavon and Wang 2009).  Second, the model
(\ref{a4}) should be constrained by local gravity experiments to avoid violation of weak equivalence principle.  It is well-known that these constraints are satisfied when $\omega>>1$ (Will 1993; Will 2055).  In the model
(\ref{a4}) this translates into $\alpha<<1$.  This means that the theory can pass local tests if interaction of the scalar
field $\varphi$ with matter fields is sufficiently small.  We will
return to this issue later.\\
We apply the field equations (\ref{a7}) and (\ref{a8}) to a spatially flat Friedmann-Robertson-Walker spacetime
\begin{equation}
ds^2=-dt^2+a^2(t)(dx^2+dy^2+dz^2)
\label{a10}\end{equation}
with $a(t)$ being the scale factor.  To do this, we take the matter system to be a pressureless perfect fluid (dust) with energy
density $\rho_m$.  In this case, the gravitational equations (\ref{a7}) give
\begin{equation}
3\frac{\dot{a}^2}{a^2}=k(\rho_{m}+\rho_{\varphi})
\label{a11}\end{equation}
\begin{equation}
2\frac{\ddot{a}}{a}+\frac{\dot{a}^2}{a^2}=-k~p_{\varphi}
\label{a12}\end{equation}
where $k=8\pi G$, $\rho_{\varphi}=\frac{1}{2}\dot{\varphi}^2+U(\varphi)$ and $p_{\varphi}=\frac{1}{2}\dot{\varphi}^2-U(\varphi)$. We may use the
first equation to rewrite the second one as
\begin{equation}
\frac{\ddot{a}}{a}+2\frac{\dot{a}^2}{a^2}=\frac{1}{2}k\rho_m+kU(\varphi)
\label{a12-1}\end{equation}
The equation (\ref{a8}) gives
\begin{equation}
\ddot{\varphi}+3\frac{\dot{a}}{a}\dot{\varphi}+\frac{dU(\varphi)}{d\varphi}=-\frac{1}{2}\alpha ~\rho_{m}
\label{a13}\end{equation}
On the other hand, the conservation
equations (\ref{1a9}) become
\begin{equation}
\dot{\rho}_{m}+3\frac{\dot{a}}{a}\rho_m=Q
\label{a14}\end{equation}
\begin{equation}
\dot{\rho}_{\varphi}+3\frac{\dot{a}}{a}(\omega_{\varphi}+1)\rho_{\varphi}=-Q
\label{a15}\end{equation}
where
\begin{equation}
Q=\frac{1}{2}\alpha~\dot{\varphi}~\rho_m
\label{a15-1}\end{equation}
and $\omega_{\varphi}=p_{\varphi}/\rho_{\varphi}$ is the equation of state parameter of the scalar field $\varphi$.
The equation (\ref{a14}) can be solved which gives the following solution
\begin{equation}
\rho_ma^3=\rho_{m0}e^{\frac{1}{2}\alpha\varphi}
\label{a16}\end{equation}
where $\rho_{m0}$ is the present matter energy density in the universe.  Now we introduce a class of solutions of
the field equations, characterized by,
\begin{equation}
\varphi=\frac{\beta}{\sqrt{k}}\ln a
\label{a17}\end{equation}
in which $\beta$ is a positive constant parameter of order of unity.
If we put the latter into (\ref{a13}), we obtain
\begin{equation}
\frac{\ddot{a}}{a}+2\frac{\dot{a}^2}{a^2}=-\frac{\sqrt{k}}{\beta}(\frac{dU(\varphi)}{\varphi}+\frac{1}{2}\alpha\rho_m)
\label{a18-1}\end{equation}
Comparing (\ref{a18-1}) with (\ref{a12-1}) leads to a consistency relation
\begin{equation}
\frac{dU(\varphi)}{d\varphi}+\sqrt{k}\beta U(\varphi)=-\frac{\sqrt{k}}{\beta}(\frac{1}{2}\beta^2+\varepsilon)\rho_{m0}e^{\frac{\sqrt{k}}{\beta}(\varepsilon-3)\varphi}
\label{a18-2}\end{equation}
in which $\varepsilon \equiv \alpha\beta /2\sqrt{k} = \beta [2(2\omega+3)]^{-\frac{1}{2}}>0$.  This consistency relation will be satisfied for an appropriate potential function $U(\varphi)$.  To find the form of this potential, we solve this first order differential equation which gives the following solution
\begin{equation}
U(\varphi)=-\gamma \rho_{m0}e^{\frac{\sqrt{k}}{\beta}(\varepsilon-3)\varphi}+Ce^{-\sqrt{k}\beta\varphi}
\label{a18-2}\end{equation}
where $\gamma=\frac{\frac{1}{2}\beta^2+\varepsilon}{\beta^2+\varepsilon-3}$ and $C$ is an integration constant.
As a consequence, the relation (\ref{a17}) is a solution of the field equations for an exponential potential of the form (\ref{a18-2}).  This double
exponential potential is similar to the potential which is used in some quintessence models (Barreiro, Copeland and Nunes 2000; Sen and Sethi 2002; Pavlov, Rubano, Sazhin and Scudellaro 2002; Demianski, Piedipalumbo, Rubano and Tortora 2005).  It is shown that this kind of potential
of the quintessence field can lead to solutions
which first enter a period of scaling through the radiation
and matter domination eras and then smoothly evolve to dominate the energy density for a wide range of initial conditions of
the field\footnote{Note that Einstein frame representation of BD models are effectively equivalent
to the so-called coupled quintessence models in which the quintessence field interacts with matter sector.} (Barreiro, Copeland and Nunes 2000).  Moreover, single exponential potentials
are popular in modified $f(R)$ gravity models (Sotiriou and Faraoni 2010; De Felice and Tsujikawa 2010).  These models are conformally equivalent to BD models with potentials which their
forms are closely related to the functional form of the $f(R)$ functions (Bisabr 2009). In that context, single exponential potentials correspond
to power law $f(R)$ gravity models (Capozziello, Cardone, Piedipalumbo and Rubano 2006; Bisabr 2010b).\\
It is interesting to note that the relation (\ref{a17}) brings the law of evolution of matter density, given by  (\ref{a16}), into
the following form
\begin{equation}
\rho_m=\rho_{m0}a^{-3+\varepsilon}
\label{a18}\end{equation}
This is similar to the rule presented by some authors for
characterizing decaying law of vacuum energy into dark matter (Wang and Meng 2005; Alcaniz and Lima 2005; Costa, Alcaniz and Maia 2008; Jesus, Santos, Alcaniz and Lima 2008; Costa and Alcaniz 2010).  It states that the scalar field $\varphi$ is constantly decaying
into the matter so that the latter will dilute
more slowly compared to its standard evolution $\rho_m \propto a^{-3}$.  Since the observational lower bound imposed by solar system experiments
on the BD parameter is $\omega >> 1$, we should have $\varepsilon<<1$ which means that evolution of matter density
has a small deviation with respect to the standard one in Einstein's gravity.\\
In the expression (\ref{a18-2}), the integration constant $C$
can be determined by noting the fact that when we set $\phi=\phi_0 \sim G^{-1}$ in the action (\ref{a1}), then $V(\phi)$ characterizes the vacuum
energy density corresponding to a cosmological constant, namely $V(\phi)=\Lambda/G$.  In this case, $\varphi=0$ and then
\begin{equation}
U(\varphi=0)=V(\phi(\varphi))\equiv\rho_{\varphi0}
\label{a18-3}\end{equation}
with $\rho_{\varphi0}$ being the vacuum energy density in the Einstein frame.  Applying the latter condition to the
relation (\ref{a18-2}) gives $C=\rho_{\varphi0}+\gamma \rho_{m0}$.  The potential function takes then the form
\begin{equation}
U(\varphi)=\gamma\rho_{m0}(e^{-\sqrt{k}\beta\varphi}-e^{\frac{\sqrt{k}}{\beta}(\varepsilon-3)\varphi})+\rho_{\varphi0}e^{-\sqrt{k}\beta\varphi}
\label{a18-4}\end{equation}
For this potential function, the Friedmann equation (\ref{a11}) becomes
\begin{equation}
\frac{H^2}{H_0^2}=\frac{3}{3-\frac{1}{2}\beta^2}[(1-\gamma)\Omega_{m0}a^{-3+\varepsilon}+ (\Omega_{\varphi 0}+\gamma\Omega_{m0})a^{-\beta^2}]
\label{a19}\end{equation}
where $\Omega_{m0}=\rho_{m0}/\rho_c$, $\Omega_{\varphi 0}=\rho_{\varphi 0}/\rho_c$ and $\rho_c=3H_0^2/k$ is the critical density.  From
the equations (\ref{a12}) and (\ref{a19}), it is straightforward to show that the deceleration parameter
\begin{equation}
q=-1-\frac{\dot{H}}{H^2}
\label{a19a}\end{equation}
 takes
the following form
\begin{equation}
q(z)=\frac{1}{2}\{(1+\frac{1}{2}\beta^2)-(3-\frac{1}{2}\beta^2)[1-\frac{1}{\gamma-(\frac{\Omega_{\varphi0}}{\Omega_{m0}}
+\gamma)(z+1)^{-3+\varepsilon+\beta^2}}]^{-1}\}
\label{a20}\end{equation}
where we have used $a(z)=(z+1)^{-1}$.   This relation gives deceleration parameter in terms of the redshift and constant parameters $\omega$ and $\beta$.  We plot
 $q(z)$ in fig.1a for $\omega=40000$.
\begin{figure}[ht]
\begin{center}
\includegraphics[width=0.45\linewidth]{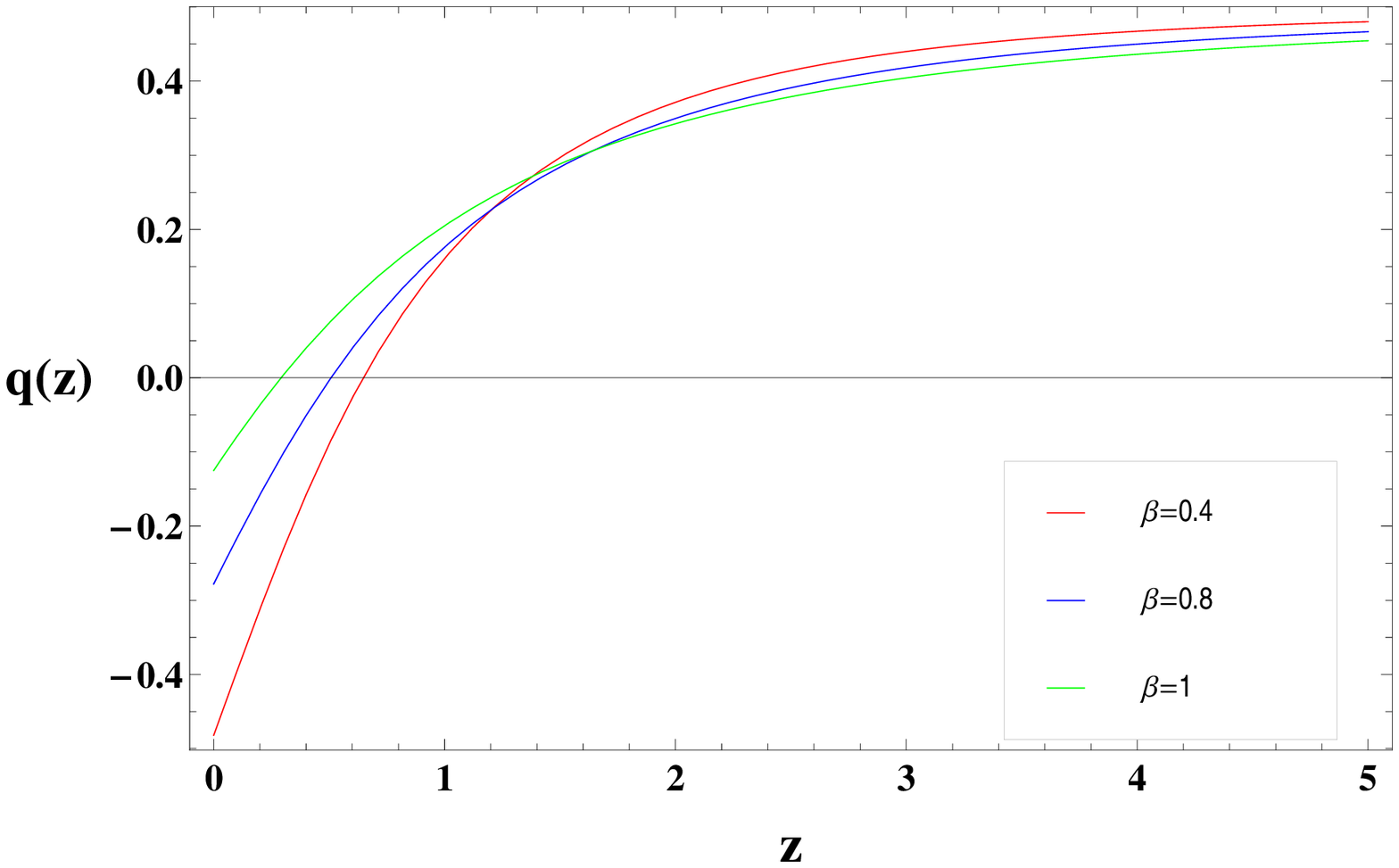}
\includegraphics[width=0.45\linewidth]{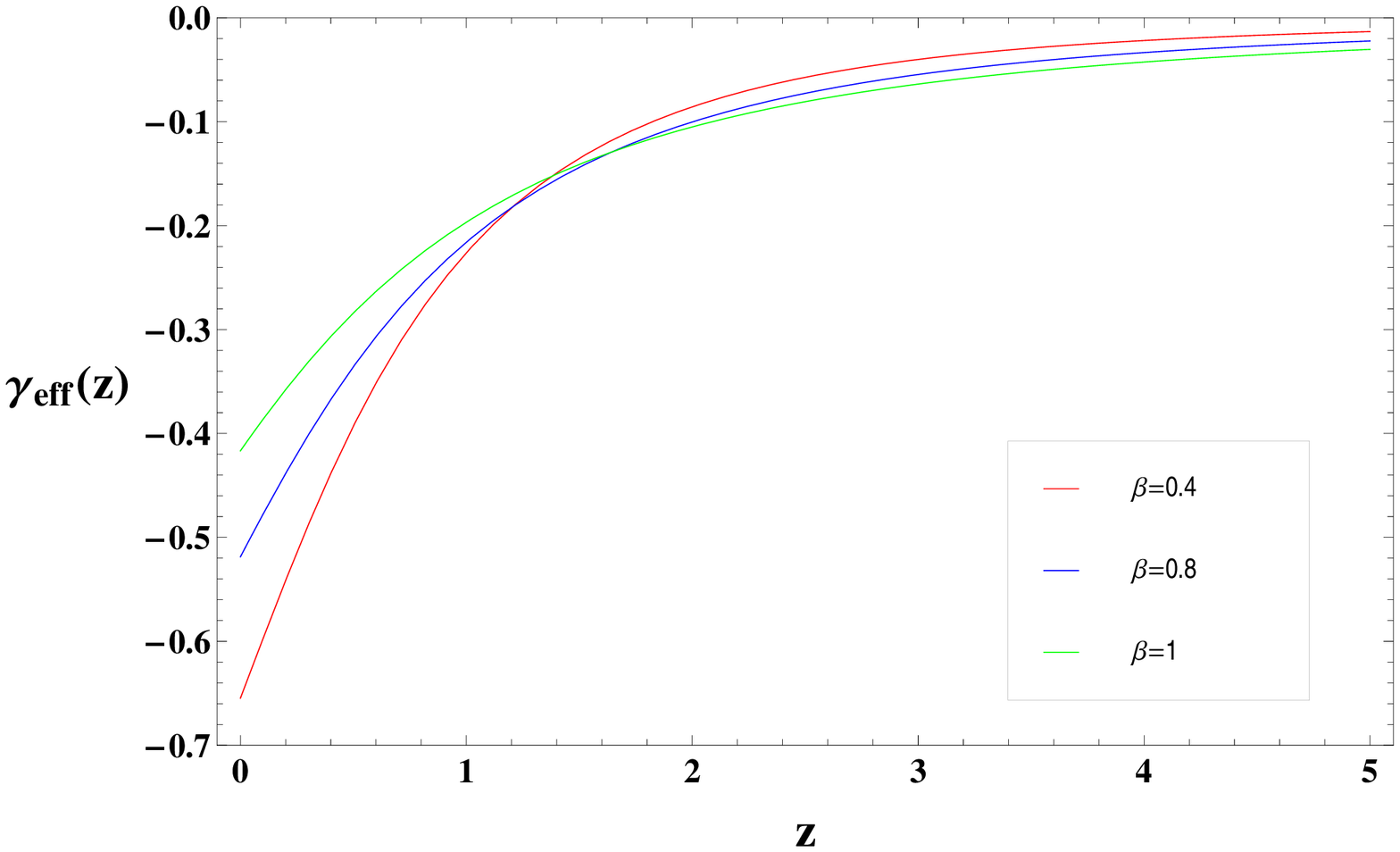}
\caption{The plot of deceleration parameter $q(z)$ (panel a) and the effective equation of state parameter (panel b), given
by (\ref{a20}) and (\ref{a21a}), for $\omega=40000$ and some values of the parameter $\beta$.  We have set
$\frac{\Omega_{m0}}{\Omega_{\varphi 0}}=\frac{3}{7}$.  The values $\beta=0.4, 0.8, 1$ respectively correspond to $\varepsilon=0.0010, 0.0020, 0.0025$ .}
\end{center}
\end{figure}
The figure indicates that accelerating expansion is possible for positive and large BD parameter ($\omega>>1$) consistent with observations. Moreover, it shows that the signature flip of $q$ corresponding to transition from decelerating to accelerating expansion takes place in
recent times as experimentally demonstrated by (Riess 2001). \\ We can also
obtain the effective equation of state parameter, defined by,
\begin{equation}
\gamma_{eff}=\frac{p_{\varphi}}{\rho_{\varphi}+\rho_{m}}
\label{a20a}\end{equation}
Combining the latter with (\ref{a11}), (\ref{a12}) and (\ref{a19a}) leads to
\begin{equation}
\gamma_{eff}=\frac{1}{3}(2q-1)
\label{a21a}\end{equation}
Evolution of $\gamma_{eff}$ with respect to the redshift is plotted in fig.1b.  The figure shows that the BD scalar field does not appear as a
phantom field in the Einstein frame since the curves do not cross the boundary $\gamma_{eff}=-1$.
\section{The Cosmic Coincidence}
One of the important features of the cosmological constant problem
is the present coincidence between dark energy and dark matter
energy densities (Steinhardt 1997; Zlatev, Wang and Steinhardt 1999).  There is a class of models in which
this observation is related to some kinds of interaction between the
two components (Zimdahl, Pavon and Chimento 2001; Zimdahl and Pavon 2003; Campo, Herrera and Pavon 2009; Wetterich 1995; Amendola 2000; Tocchini-Valentini and Amendola 2002; Boehmer, Caldera-Cabral, Lazkoz and Maartens 2008).  In these models the two components are not
separately conserved and there is a flow of energy from dark
energy to dark matter or vice versa. In this sense,
dark energy and dark matter energy densities may
have the same scaling at late times due to the interaction, although they decrease with the
expansion of the universe at different rates.
 The important task in this context is
to find a constant ratio of dark energy to dark matter energy densities
for an appropriate interaction term.  Despite the fact that this approach seems to
be promising, there is not still a
compelling form of interaction which is introduced by a
fundamental theory.  Therefore one usually uses different
interaction terms and tries to adapt them with recent observations.
In BD theory presented in Einstein frame, there is
a fixed interaction between the scalar field and matter sector induced by conformal transformation.  The strategy that we are going to
pursue in this section is to use this interaction term for finding evolution of the ratio $r\equiv \rho_{m}/\rho_{\varphi}$.\\
To do this, we first write the expression (\ref{a18}) in the following form
\begin{equation}
\rho_m=\frac{3H_0^2}{k}\Omega_{m0}a^{-3+\varepsilon}
\label{a21}\end{equation}
Then, we use (\ref{a17}) to write
\begin{equation}
\rho_{\varphi}=\frac{1}{2}\dot{\varphi}^2+U(\varphi)=\frac{H_0^2}{k}\{\frac{1}{2}\beta^2\frac{H^2}{H_0^2}+3\gamma \Omega_{m0}(a^{-\beta^2}-a^{-3+\varepsilon}
)+3\Omega_{\varphi 0}a^{-\beta^2}\}
\label{a22}\end{equation}
We may combine the latter with (\ref{a19}) to obtain
\begin{equation}
\rho_{\varphi}=\frac{H_0^2}{k}\{3\Omega_{m0} \frac{\beta^2-6\gamma}{6-\beta^2}a^{-3+\varepsilon} +\frac{18}{6-\beta^2}
(\gamma \Omega_{m0}+\Omega_{\varphi 0})a^{-\beta^2}\}
\label{a23}\end{equation}
Finally, expressions (\ref{a21}) and (\ref{a23}) give the ratio of energy densities
\begin{equation}
r(z)=3\Omega_{m0}(z+1)^{3-\varepsilon}\{3\Omega_{m0}\frac{\beta^2-6\gamma}{6-\beta^2}(z+1)^{3-\varepsilon}+\frac{18}{6-\beta^2}
(\gamma \Omega_{m0}+\Omega_{\varphi 0})(z+1)^{\beta^2}\}^{-1}
\label{a24}\end{equation}
We plot $r(z)$ in fig.2.  The figure indicates that the ratio $r(z)$ varies during expansion of the universe and tends then to some constant of order unity in recent times.  This behavior seems to be independent of the value attributed to the parameter $\beta$.  However, the latter affect the variation
of $r$ at large redshifts.  When $\beta$ increases, which corresponds to increasing of $\varepsilon$, variation of $r(z)$ decreases.  This implies that the effect
of interaction of the two energy components is to make the ratio $r$ have more smooth variations at large redshifts.
\begin{figure}[ht]
\begin{center}
\includegraphics[width=0.6\linewidth]{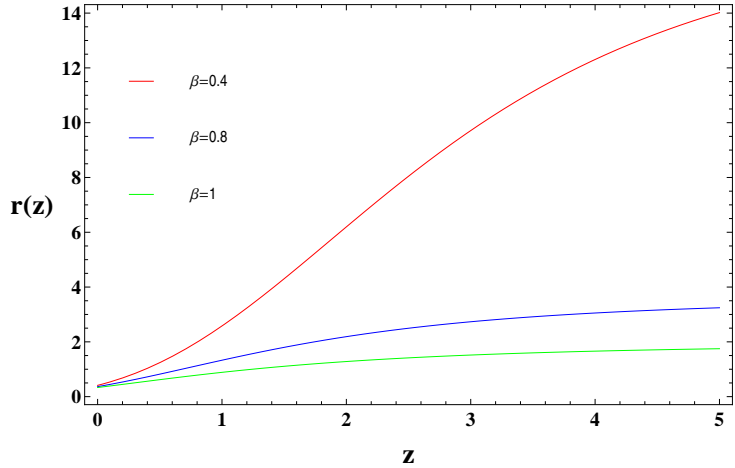}
\caption{The plot of $r(z)$ for $\omega=40000$ and some values of the parameter $\beta$.  We have set
$\frac{\Omega_{m0}}{\Omega_{\varphi 0}}=\frac{3}{7}$. }
\end{center}
\end{figure}

\section{Conclusion}
We have considered the possibility that a self-interacting Brans-Dicke field accounts for the
accelerated expansion of the Universe in Einstein conformal frame.  Our analysis is limited to a class
of solutions characterized by (\ref{a17}) which is given in terms of a double exponential potential of the form (\ref{a18-2}).
The first important feature of the solutions is that they provide
a late-time accelerating expansion of the universe for large values of the BD parameter $\omega$.  At the same time, these solutions
modify the evolution of matter density to (\ref{a18}) which is the simplest possible way of stating that
the matter dilution is attenuated due to its interaction with the scalar field $\varphi$.  This evolution law indicates that the deviation from the standard evolution is characterized
by a positive constant parameter $\varepsilon$ which quantifies the decay rate.  Accelerating
expansion is possible for $\varepsilon<<1$ (or $\omega>>1$) in accord with local gravity tests. \\
Another important feature of the model presented here is that it simultaneously solves the coincidence problem.  As previously stated, the interaction of the BD field
$\varphi$ with dark sector has been recently taken as a natural guidance for addressing coincidence problem by some authors.  However, in absence of an interaction or coupling term based on a fundamental theory, most of the current investigations have been limited to a phenomenological level.  In our analysis, this interaction term is given by the conformal transformation.  We have investigated the consequences of this interaction
 term and derived a relation giving the evolution of the ratio of energy densities
$r$.  We have found that in spite of the fact that the ratio can be variable during expansion of the universe in the past, it tends to some constant value of order unity at late-time thus solving the coincidence problem.

\end{document}